\documentclass[conference]{IEEEtran}
\IEEEoverridecommandlockouts
\usepackage{cite}
\usepackage{amsmath,amssymb,amsfonts}
\usepackage{algorithmic}
\usepackage{graphicx}
\usepackage{textcomp}
\usepackage{xcolor}
\usepackage{tabularx}
\usepackage{booktabs}
\usepackage{multirow}
\usepackage{makecell}
\def\BibTeX{{\rm B\kern-.05em{\sc i\kern-.025em b}\kern-.08em
    T\kern-.1667em\lower.7ex\hbox{E}\kern-.125emX}}
\begin{document}
\title{Open-Set Source Tracing as Compositional Factors via Structured Prototypes
}

\author{\IEEEauthorblockN{Santiago Rubio}
\IEEEauthorblockA{\textit{ViVoLab, I3A} \\
\textit{University of Zaragoza}\\
Zaragoza, Spain \\
s.rubio@unizar.es}
\and
\IEEEauthorblockN{Antonio Almudévar}
\IEEEauthorblockA{\textit{ViVoLab, I3A} \\
\textit{University of Zaragoza}\\
Zaragoza, Spain \\
almudevar@unizar.es}
\and
\IEEEauthorblockN{Antonio Miguel}
\IEEEauthorblockA{\textit{ViVoLab, I3A} \\
\textit{University of Zaragoza}\\
Zaragoza, Spain \\
amiguel@unizar.es}
\and
\IEEEauthorblockN{Eduardo Lleida}
\IEEEauthorblockA{\textit{ViVoLab, I3A} \\
\textit{University of Zaragoza}\\
Zaragoza, Spain \\
lleida@unizar.es}
\and
\IEEEauthorblockN{Alfonso Ortega}
\IEEEauthorblockA{\textit{ViVoLab, I3A} \\
\textit{University of Zaragoza}\\
Zaragoza, Spain \\
ortega@unizar.es}
}

\maketitle

\begin{abstract}
Recent research expands beyond binary anti-spoofing with the emergence of Source Tracing, the task of identifying the specific generative origins of synthetic speech. However, current research often equates a ``source'' with its generative architecture. We propose redefining a source as a compositional tuple of Architecture, Training Data, and other training factors affecting the generated speech.
We propose a framework using Structured Orthonormal Prototypes to minimize class overlap and intra-class variance. Our Subspace Partitioning strategy splits the embedding into architecture and data subspaces, while a residual subspace captures stochastic variability, enabling ``compositional generalization'' for novel factor combinations. This approach improves performance for partially seen sources and maintains robustness in fully open-set scenarios. MLAAD evaluations for Few-Shot open-set Identification show our approach significantly outperforms angular-margin baselines.
\end{abstract}

\begin{IEEEkeywords}
audio deepfake detection, source tracing, 
prototype learning, open-set attribution
\end{IEEEkeywords}

\section{Introduction}

High-fidelity neural audio synthesizers are redefining audio forensics. The traditional binary spoofing detection task, formalized by benchmarks like ASVspoof~\cite{ASVspoof15,ASVspoof17,ASVspoof19,ASVspoof21,ASVspoof24} and ADD~\cite{ADD2022,ADD2023}, is increasingly insufficient given the rapid expansion and diversity of modern generative models. This heterogeneity has driven the emergence of Source Tracing~\cite{klein24_interspeech}, which seeks to identify the specific model that generated a spoofed sample. Recent datasets, including MLAAD~\cite{10650962}, STOPA~\cite{firc25_interspeech}, and CodecFake+~\cite{chen2025codecfake+}, support this direction. However, despite emerging resources, the task lacks a formal problem statement.

Current literature predominantly equates a ``source'' with its generative architecture, either by decomposing sources into purely algorithmic components~\cite{falez25_interspeech,chen25j_interspeech,chhibber26_odyssey} or by adapting speaker recognition paradigms to treat these architectural labels as monolithic identities~\cite{negroni25_interspeech}. Even open-set approaches~\cite{klein25_interspeech} prioritize out-of-distribution detection scores over characterizing the source's constitutive factors. We argue this architectural focus is forensically insufficient. Drawing a parallel to the ``Nature vs. Nurture'' distinction~\cite{7b3f0a2f-4c9b-34db-b425-bd05e12a6d9b}, a generative model is a stochastic realization shaped by its architectural inductive biases (akin to ``Nature'', $\mathcal{A}$) and the empirical distribution of its training data (akin to ``Nurture'', $\mathcal{D}$), plus the residual training configuration $\mathcal{H}$ that ``\textit{fingerprints}'' it beyond the two. Consequently, we redefine the source space formally as the tuple $\mathcal{S}=(\mathcal{A},\mathcal{D},\mathcal{H})$, where $\mathcal{H}$ captures the training-stage variability (hyperparameters, optimization) not explained by $\mathcal{A}$ and $\mathcal{D}$.

However, translating this into discriminative features remains challenging. Standard Deep Metric Learning relies on contrastive~\cite{kulkarni25_interspeech} or angular margin losses~\cite{koutsianos25_interspeech} that compress the entire source into a single embedding. While recent work has attempted to deconstruct modular architectural components~\cite{firc25_interspeech}, these approaches lack explicit constraints to separate algorithmic features from data-driven traits. We hypothesize this entanglement is the primary bottleneck for open-set attribution: without separating $\mathcal{A}$ from $\mathcal{D}$, forensic models cannot generalize to known architecture trained on unseen data (or vice versa).

To bridge this gap, we propose a prototype-based framework using geometric constraints to separate architectural and data-driven factors. Inspired by advances in interpretable deep learning~\cite{almudevar24_interspeech,electronics10070850}, we employ explicit prototypes to impose structure on the metric space rather than relying on spontaneous cluster emergence. Our framework introduces two key strategies: 
(i) \textbf{Structured Orthonormal Prototypes}, frozen to enforce geometric regularity, and (ii) \textbf{Subspace Partitioning} that splits the embedding into separate architecture and data-driven subspaces, plus a residual subspace $\mathcal{Z}_{\mathcal{R}}$ that captures the residual training configuration $\mathcal{H}$ together with any other unlabeled variability. This factorization enables \textit{compositional generalization}: the model can attribute audio from novel architecture and data combinations. This improves performance for unseen pairings of previously known architectures and datasets, with more limited but consistent gains in fully open-set scenarios where one or both generative factors are completely new.

\begin{figure*}[t]
    \centering
    \includegraphics[width=0.9\textwidth]{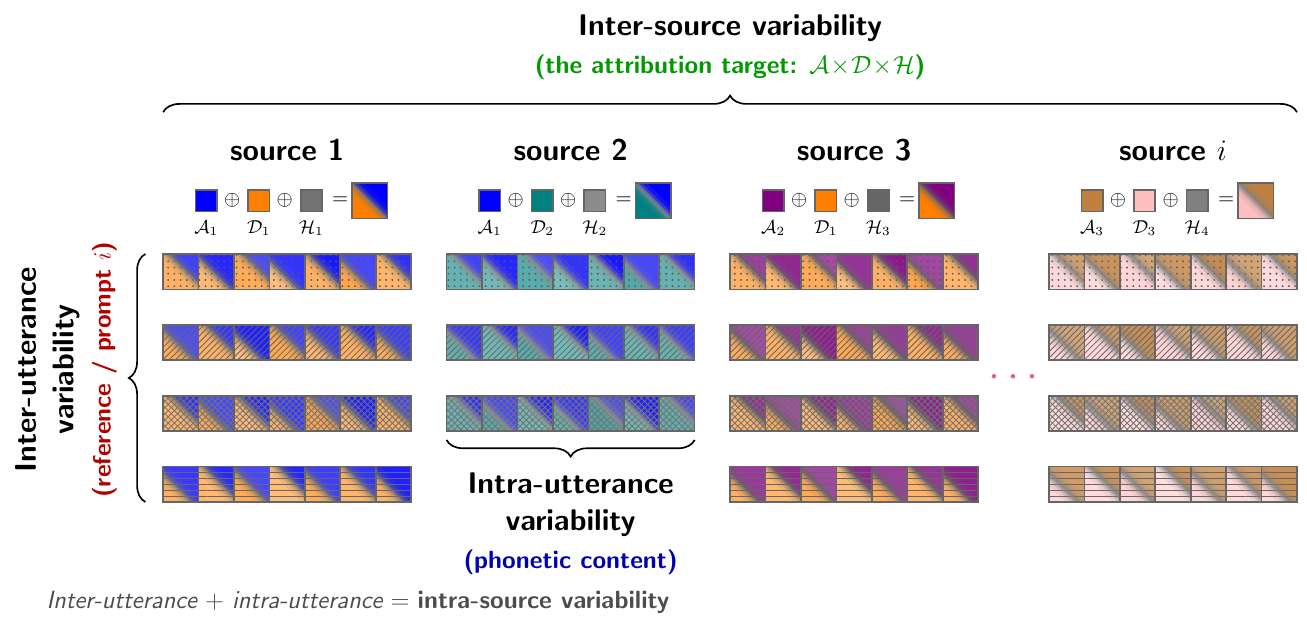}
    \vspace{-6pt}
    \caption{Variability in synthetic speech attribution. A source $\mathcal{S}=(\mathcal{A},\mathcal{D},\mathcal{H})$ is read as a composition of an architecture factor ($\mathcal{A}$, upper-right triangle), a training-data factor ($\mathcal{D}$, lower-left triangle) and a residual ($\mathcal{H}$, the grey band blurring their boundary); sources sharing a factor share the corresponding color, and the $\mathcal{A}$/$\mathcal{D}$ boundary shifts from tile to tile as one factor dominates. \emph{Inter-source} variability (across columns) is the attribution target, whereas \emph{inter-utterance} variability (across rows: reference speaker or prompt, shown as a distinct per-row texture) and \emph{intra-utterance} variability (within a row: phonetic content) constitute the intra-source nuisance to be compensated for attribution.}
    \label{fig:variability}
\end{figure*}

\section{Methodology}
\label{sec:Methodology}

\subsection{Variability in Synthetic Speech Attribution}
\label{sec:variability}
A synthetic utterance blends several kinds of variation at once, and not all of them are equally useful for attribution: some reveal which model produced it, while most only reflect how that particular utterance was rendered. A source tracing system must therefore separate the factors that \emph{define} the generative source from the variability that merely \emph{renders} each observation. We organize this variability into the hierarchy of Figure~\ref{fig:variability}. At the top level, \emph{inter-source} variability distinguishes one generative model from another and is the attribution target: each source is a composition of an architecture factor $\mathcal{A}$, a training-data factor $\mathcal{D}$, and a residual $\mathcal{H}$, so that sources sharing an architecture or a dataset share part of their identity. Beneath it lies \emph{intra-source} variability, the variation a single source produces across its own outputs, which we split into two levels: \emph{inter-utterance} variability, induced by the reference speaker or prompt that conditions a given synthesis, and \emph{intra-utterance} variability, induced by the phonetic content within an utterance. Only the inter-source level carries useful information for source tracing tasks; the intra-source levels must be compensated for robust attribution. This decomposition mirrors, while inverting, the variability compensation long studied in speaker verification: there the speaker is the target and channel or content the nuisance; here the roles invert, with the generative source becoming the target, while the inference-time inputs that condition it, such as the prompt and reference speaker, become the nuisance~\cite{negroni25_interspeech,dao2026speaker}.
\subsection{Problem Formulation}
\label{sec:problem_formulation}

We redefine the identity of a synthetic sample as a result of a compositional process, distinguishing two levels of variability. The first level defines the generative model itself and is fixed at training time: its architecture $a \in \mathcal{A}$, its training data $d \in \mathcal{D}$, and its training configuration $h \in \mathcal{H}$, i.e., the hyperparameters, seed, and optimization stochasticity that leave a reproducible fingerprint even when $\mathcal{A}$ and $\mathcal{D}$ are held fixed, together with the non-linear $\mathcal{A}\times\mathcal{D}$ interactions that marginal modeling cannot express. We thus define a source as the tuple $s=(a,d,h)$. A second level of variability arises at inference time from the \emph{inference variables} $i \in \mathcal{I}$ that render a particular synthesis, such as the reference speaker, the phonetic content, or the prompt. A synthetic utterance is then a realization $x = G_s(i,\varepsilon)$ of source $s$ under inference condition $i$ and generative noise $\varepsilon$. Source tracing seeks to recover $s$ while remaining invariant to $i$. Because $i$ changes across utterances of a single source, it is a nuisance to be marginalized rather than a constituent of the source: the reference speaker in modern generative speech models, in particular, is an inference-time factor and not an additional identity axis. Some factors of both $i$ and $h$, such as the sampling temperature or the optimization hyperparameters, are not commonly recorded as labels, and are therefore accommodated by the residual subspace $\mathcal{Z}_{\mathcal{R}}$ introduced below.

\begin{table}[!tb]
\centering
\caption{Scenarios for Synthetic Speech Attribution}
\label{tab:scenarios}
\small% Keeps text readable but gives the table more room
\begin{tabularx}{\columnwidth}{@{} l >{\raggedright\arraybackslash}p{2.6cm} >{\raggedright\arraybackslash\footnotesize}X
@{}}
\toprule
\textbf{Scenario} & \textbf{ Formal Definition} & \textbf{ Description} \\ \midrule
\makecell[tl]{Fully Closed \\ {\scriptsize (IID)}} & $(a, d) \in \mathcal{T}_{\text{seen}}$ & Both arch. and data seen together in train. \\ \addlinespace
\makecell[tl]{Compositional \\ {\scriptsize (OOD)}} & $a \in \mathcal{A}_{\text{tr}}, d \in \mathcal{D}_{\text{tr}},$ $(a, d) \notin \mathcal{T}_{\text{seen}}$ & Factors seen separately, but this pairing is novel. \\ \addlinespace
\makecell[tl]{Partial Open \\ {\scriptsize (OOD)}} & $(a \in \mathcal{A}_{\text{tr}}, d \notin \mathcal{D}_{\text{tr}}) \lor$ $(a \notin \mathcal{A}_{\text{tr}}, d \in \mathcal{D}_{\text{tr}})$ & One factor is seen, while the other is new. \\ \addlinespace
\makecell[tl]{Fully Open \\ {\scriptsize (OOD)}}
 & $a \notin \mathcal{A}_{\text{tr}}, d \notin \mathcal{D}_{\text{tr}}$ & Both arch. and data are novel to the model. \\ \bottomrule
\end{tabularx}
\end{table}

\begin{figure*}[t]
    \centering
    \includegraphics[width=0.95\linewidth]{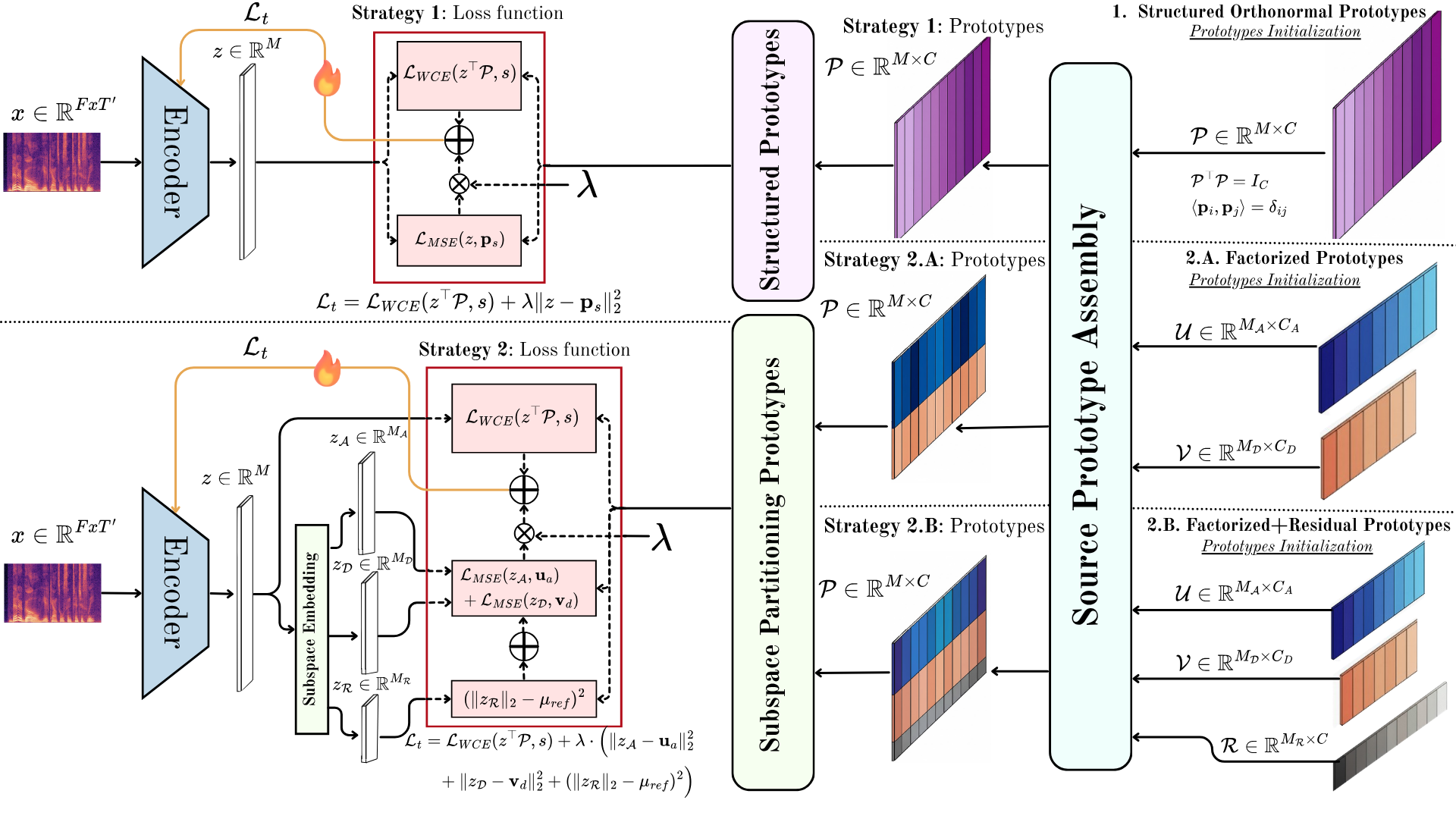}
    \caption{Overview of the Proposed Source Tracing Framework. Top (Strategy 1): Structured Orthonormal Prototypes constrained by $\mathcal{L}_{WCE}$ and $\mathcal{L}_{MSE}$. Bottom (Strategy 2): Our proposed Subspace Partitioning Framework. The embedding is split into factors: Architecture ($z_{\mathcal{A}}$, blue), Training Data ($z_{\mathcal{D}}$, orange), and Residual Information ($z_{\mathcal{R}}$, grey).}
    \label{fig:schema}
\end{figure*}
Therefore, the source tracing task is transformed from monolithic classification into a factorized attribution problem. Let $\mathcal{A}_{\text{tr}}$ and $\mathcal{D}_{\text{tr}}$ be the architectures and datasets seen in training. Given the set of pairings seen during training ${\mathcal{T}_{\text{seen}} \subset \mathcal{A}_{\text{tr}} \times \mathcal{D}_{\text{tr}}}$, we define the task through four mutually exclusive scenarios covering all configurations, as detailed in Table~\ref{tab:scenarios}.

Our objective is to achieve \textbf{compositional generalization}: attributing audio from novel combinations $(a, d)$ even for unseen pairs. We achieve this by anchoring the factorized subspaces $\mathcal{Z}_{\mathcal{A}}$ and $\mathcal{Z}_{\mathcal{D}}$ to explicit prototypes built from the architecture and dataset labels available in metadata, capturing the marginal structure of $\mathcal{A}$ and $\mathcal{D}$. These are the primary components of source identity, but real-world generation is also shaped by factors that carry no such labels, namely the training residual $h$, the non-linear $\mathcal{A}\times\mathcal{D}$ interactions, and any unlabeled inference-time variability, which therefore admit no explicit target to align to.
Consequently, we introduce a dedicated, prototype-free residual subspace $\mathcal{Z}_{\mathcal{R}}$ that acts as a \textbf{residual manifold}: it absorbs these factors jointly into a latent space where they can be effectively combined, so that the explicit factorization operates solely on the labeled factors $\mathcal{A}$ and $\mathcal{D}$.

\subsection{Learning Strategies}

To structure the latent space for optimal few-shot performance, we investigate three progressive learning strategies over the encoder's output embedding $z \in \mathbb{R}^M$, where $M$ denotes the embedding dimensionality (Figure~\ref{fig:schema}).

\textbf{Baseline: Angular Margin (ArcFace).} As a standard baseline, we employ ArcFace~\cite{Deng_2022}, which learns a weight matrix $W \in \mathbb{R}^{M \times C}$ by maximizing the geodesic margin on the hypersphere, where $C$ is the total number of seen classes. Denoting by $\theta_{s,c}$ the angle between the embedding $z$ and the class weight $W_c$, the objective is:
\begin{equation}
\mathcal{L}_{ArcFace} = -\log \frac{e^{\,s\,\cos(\theta_{s,s} + m)}}{e^{\,s\,\cos(\theta_{s,s} + m)} + \sum_{c \neq s} e^{\,s\,\cos\theta_{s,c}}}
\end{equation}
where $m$ is the additive angular margin and $s$ the scale factor. This imposes no explicit structure on how generative factors are organized, which may limit open-set generalization.

\textbf{Strategy 1: Orthonormal Prototypes.} To impose maximal inter-class separability, we replace the learnable weights with a fixed Prototype Matrix $\mathcal{P} \in \mathbb{R}^{M \times C}$. Since $C \le M$, we initialize $\mathcal{P}$ such that its columns form an orthonormal set (for $C > M$, exact orthogonality requires increasing $M$ or using near-orthogonal spherical codes~\cite{johnson1984extensions}).
We consider all sources as mutually independent and therefore priors are selected maximally distant in the angular space. The encoder is optimized to align the embedding $z$ with its assigned basis vector $\mathbf{p}_s \in \mathcal{P}$ using a constrained objective:
\begin{equation}\mathcal{L}_{Ortho} = \mathcal{L}_{WCE}(z^\top \mathcal{P}, s) + \lambda \lVert z - \mathbf{p}_{s} \rVert_2^2
\end{equation}
where $\mathcal{L}_{WCE}$ is a Weighted Cross-Entropy loss with class weights inversely proportional to source frequency. The MSE term encourages embeddings to remain close to their assigned prototypes, acting as a soft regularizer that balances angular discrimination with geometric stability.

\textbf{Strategy 2: Subspace Partitioning.} Strategy 1's strict orthogonality assumes equidistant sources, ignoring shared characteristics between models using the same architecture or training data. To capture this factorial structure, we partition the embedding space $\mathbb{R}^M$ into a structured set of orthogonal subspaces.

\textbf{Strategy 2.A: Factorized Prototypes.} We decompose the dimension $M$ into $M_{\mathcal{A}} + M_{\mathcal{D}}$. Let $\mathcal{U} = \{\mathbf{u}_a\}_{a=1}^{C_A}\subseteq \mathbb{R}^{M_{\mathcal{A}}}$ and $\mathcal{V} = \{\mathbf{v}_d\}_{d=1}^{C_D}\subseteq \mathbb{R}^{M_{\mathcal{D}}}$ be orthonormal sets, where $\mathbf{u}_a$ and $\mathbf{v}_d$ represent specific instances among the $C_A$ architecture and $C_D$ training data factors, respectively.
For a source $s$ defined by the tuple $(\mathcal{A}_a, \mathcal{D}_d)$, we build a composite prototype $\mathbf{p}_s$ via concatenation: $\mathbf{p}_s = \mathbf{u}_a \oplus \mathbf{v}_d$.  The full prototype matrix $\mathcal{P} \in \mathbb{R}^{M \times C}$ is thus assembled column-wise from all such composite prototypes. Unlike Strategy 1, these prototypes are not mutually orthogonal; sources sharing an architecture $a$ will have a non-zero dot product due to the shared component $\mathbf{u}_a$. The embedding is partitioned as $z = z_{\mathcal{A}} \oplus z_{\mathcal{D}}$. The loss minimizes the alignment error of each factor to its prototype:
\begin{equation}
\begin{aligned}
\mathcal{L}_{Fact}& = \mathcal{L}_{WCE}(z^\top \mathcal{P}, s)\\ &+ \lambda \left( \lVert z_{\mathcal{A}} - \mathbf{u}_{a} \rVert_2^2 + \lVert z_{\mathcal{D}} - \mathbf{v}_{d} \rVert_2^2 \right)
\end{aligned}
\end{equation}
Since $\mathbf{u}_a$ and $\mathbf{v}_d$ are unit vectors, this objective implicitly regularizes the norm of $\lVert z_{\mathcal{A}}\rVert$ and $\lVert z_{\mathcal{D}}\rVert$. %towards 1.

\textbf{Strategy 2.B: Residual Modeling.} To capture the residual factors of Section~\ref{sec:problem_formulation}, we extend the partition with a residual subspace: $z = z_{\mathcal{A}} \oplus z_{\mathcal{D}} \oplus z_{\mathcal{R}}$, with $z_{\mathcal{R}} \in \mathbb{R}^{M_{\mathcal{R}}}$.
Since $z_{\mathcal{R}}$ corresponds to non-deterministic factors, it lacks a fixed target prototype. Minimizing a standard MSE would tend to collapse $z_{\mathcal{R}}$ toward zero, while leaving it unconstrained allows unbounded growth. To address this, we introduce an \textbf{Energy Constraint} inspired by the soft norm regularization of Ring Loss~\cite{Zheng_2018_CVPR} adapted to our factorized setting:

\begin{equation}
\mathcal{L}_{Total} = \mathcal{L}_{Fact} + \lambda \left( \lVert z_{\mathcal{R}}\rVert_2 - \mu_{ref} \right)^2
\end{equation}
where the reference magnitude $\mu_{ref} = \frac{1}{2}(\lVert\mathbf{u}_a\rVert_2 + \lVert\mathbf{v}_d\rVert_2)=1$ discourages both representation collapse~\cite{bardes2022vicreg} and norm dominance, so that $z_{\mathcal{R}}$ models bounded uncertainty without dominating $\mathcal{Z}_\mathcal{A}$ and $\mathcal{Z}_\mathcal{D}$. We target a fixed magnitude rather than an upper bound so that every source receives a comparable residual energy, letting $\mathcal{Z}_{\mathcal{R}}$ encode the \emph{direction} of the residual rather than its magnitude, which suits the cosine-similarity attribution.

Because $\mathbf{u}_a$ and $\mathbf{v}_d$ are fixed, the alignment terms drive $z_{\mathcal{A}}$ and $z_{\mathcal{D}}$ toward the same targets for every utterance of a source; variability tied to $i$ is therefore absorbed by the residual subspace $\mathcal{Z}_{\mathcal{R}}$ rather than by the factor subspaces.

\section{Experimental Setup}

\subsection{Dataset}
We use the MLAAD dataset~\cite{10650962}, a standard Source Tracing benchmark. For synthesis categorization, we adopt the creators' taxonomy, using the original metadata for model architectures and training data without re-annotation.
Our explicit factors are drawn from the architecture and training-data metadata. Speaker identity is one of the inference variables $i \in \mathcal{I}$ of Section~\ref{sec:problem_formulation} (an inter-utterance factor); since MLAAD's speaker labels are inconsistent across sources, we do not use them to compensate for this variability. As bonafide speech, we employ the M-AILABS dataset~\cite{Pratap2020MLSAL}. Since it was used to generate a significant portion of MLAAD's synthetic samples, this minimizes acoustic mismatch, ensuring the model focuses on synthesis artifacts rather than dataset-specific attributes.

The training set contains 24 sources, from 19 architectures and 11 datasets (Table~\ref{tab:dataset_stats}). For dataset labels, we include language information (e.g., ljspeech\_en, css10\_de), except for inherently multilingual sources where only the dataset is specified (e.g., multi-dataset). Sources with undocumented training data receive independent unknown labels. 
\begin{table}[!ht]
\centering
\caption{Dataset composition. Parentheses denote unseen (OOD) 
entities absent from the Training set.}
\label{tab:dataset_stats}
\resizebox{\columnwidth}{!}{% Ajusta la tabla al ancho de la columna
\begin{tabular}{ll c ccc}
\toprule
 & & \textbf{Samples} & \multicolumn{3}{c}{\textbf{MLAAD Diversity}} \\
\cmidrule(lr){4-6}
\textbf{Split} & \textbf{Subset} & \textbf{(Count)} & \textbf{Source} & \textbf{Arch.} & \textbf{Data.} \\
\midrule
\textbf{Train} & MLAAD (Spoof) & 11,100 & 24 & 19 & 11 \\
 & M-AILABS (Bonafide) & 3,508 & - & - & - \\
\midrule
\textbf{Val} & MLAAD (Spoof) & 12,000 & 25 \textbf{(17)} & 18 \textbf{(8)} & 11 (2) \\
 & M-AILABS (Bonafide) & 3,792 & - & - & - \\
\midrule
\textbf{Eval} & MLAAD (Spoof) & 33,900 & 64 \textbf{(43)} & 41 \textbf{(24)} & 21 (10) \\
 & M-AILABS (Bonafide) & 10,716 & - & - & - \\
\bottomrule
\end{tabular}
}
\end{table}

This composition exhibits natural factor reuse, pairing architectures with multiple datasets and vice versa. To rigorously evaluate generalization, we maintain an \textbf{Out-Of-Distribution (OOD)} setting where many evaluation sources, architectures, and datasets remain unseen during training (Table~\ref{tab:dataset_stats}). We evaluate five scenarios following Table~\ref{tab:scenarios}, splitting \textit{Partial Open} into \textit{Unseen Architecture} and \textit{Unseen Dataset}.

\subsection{Architecture and Training Details}

We use a ResNet18~\cite{he2015deepresiduallearningimage} over Log-Mel Spectrograms ($n_{\text{mels}}=80$, $n_{\text{fft}}=512$, hop $=160$) to produce an embedding $z \in \mathbb{R}^{M}$ ($M=128$). All strategies share this encoder and front-end, so that the comparison isolates the effect of the metric-space structure under study, with ArcFace trained on the same encoder as the angular-margin reference. Audio is resampled to 16\,kHz and preprocessed with power normalization and random silence trimming to prevent shortcuts based on loudness or padding artifacts. We trained for 25 epochs with batch size 32, using Adam and Cosine Annealing scheduling. For Strategy~2.A, $M_{\mathcal{A}} = M_{\mathcal{D}}= 64$; for Strategy~2.B, $M_{\mathcal{A}}= M_{\mathcal{D}} = 48$ and $M_{\mathcal{R}} = 32$.

\begin{table*}[t]
\centering
\caption{\textbf{Global Source Attribution Performance (F1-Macro).} Comparison of Closed-Set (Seen IID) and Few-Shot Open-Set attribution. We report \textbf{All vs All} at $K=1,3,5$. To analyze the source of improvements at $K=5$, we report the decomposition into Seen and Unseen sources. \textbf{Bold} indicates best performance per column.}
\label{tab:allkshot_f1_macro_cosine}
\setlength{\tabcolsep}{3pt}
\renewcommand{\arraystretch}{1.1}
\begin{tabular}{l c c ccc cc c}
\toprule
\multirow{4.5}{*}{\textbf{Method}} & \multirow{4.5}{*}{\textbf{MSE ($\lambda$)}} & \textbf{\small Closed-Set} & \multicolumn{6}{c}{\small \textbf{Few-Shot Open-Set}} \\ 
\cmidrule(lr){3-3} \cmidrule(lr){4-9}
 & & \multirow{2}{*}{\textbf{\begin{tabular}{@{}c@{}}Seen\\(IID)\end{tabular}}} & \multicolumn{3}{c}{\textbf{All vs All}} & \footnotesize{\textbf{Seen vs All}} & \footnotesize{\textbf{Unseen vs All}} & \footnotesize{\textbf{Gen.}} \\
\cmidrule(lr){4-6} \cmidrule(lr){7-7} \cmidrule(lr){8-8} \cmidrule(lr){9-9}
 & & & $K=1$ & $K=3$ & $K=5$ & $K=5$ & $K=5$ & \footnotesize{\textbf{Gap} $\downarrow$} \\
\midrule

% --- BASELINE ---
\textbf{Baseline} \footnotesize{(ArcFace)} & -- & 95.11 \scriptsize{$\pm$ 0.18} & 68.01 \scriptsize{$\pm$ 0.56} & 77.49 \scriptsize{$\pm$ 0.56} & 80.11 \scriptsize{$\pm$ 0.53} & 89.17 \scriptsize{$\pm$ 0.86} & 75.47 \scriptsize{$\pm$ 0.67} & 13.70 \scriptsize{$\pm$ 1.09} \\

\midrule

% --- STRATEGY 1 ---
\multirow{4}{*}{\shortstack[l]{\textbf{Strategy 1} \\ \footnotesize{(Orthonormal} \\ \footnotesize{Prototypes)}}}
 & 0.0  & 95.20 \scriptsize{$\pm$ 0.16} & 70.18 \scriptsize{$\pm$ 0.80} & 80.09 \scriptsize{$\pm$ 0.61} & 82.70 \scriptsize{$\pm$ 0.55} & 90.68 \scriptsize{$\pm$ 0.26} & 78.61 \scriptsize{$\pm$ 0.82} & 12.07 \scriptsize{$\pm$ 0.86} \\
 & 0.05 & 95.34 \scriptsize{$\pm$ 0.05} & 70.35 \scriptsize{$\pm$ 0.83} & 80.10 \scriptsize{$\pm$ 0.51} & 82.56 \scriptsize{$\pm$ 0.47} & 91.14 \scriptsize{$\pm$ 0.11} & 78.17 \scriptsize{$\pm$ 0.71} & 12.97 \scriptsize{$\pm$ 0.72} \\
 & 0.1  & 95.31 \scriptsize{$\pm$ 0.12} & 70.52 \scriptsize{$\pm$ 0.49} & 80.15 \scriptsize{$\pm$ 0.34} & 82.55 \scriptsize{$\pm$ 0.29} & 91.22 \scriptsize{$\pm$ 0.10} & 78.11 \scriptsize{$\pm$ 0.44} & 13.11 \scriptsize{$\pm$ 0.45} \\
 & 1.0  & \textbf{95.44 \scriptsize{$\pm$ 0.05}} & 69.38 \scriptsize{$\pm$ 0.46} & 78.99 \scriptsize{$\pm$ 0.25} & 81.44 \scriptsize{$\pm$ 0.18} & 91.39 \scriptsize{$\pm$ 0.18} & 76.35 \scriptsize{$\pm$ 0.25} & 15.04 \scriptsize{$\pm$ 0.31} \\

\midrule

% --- STRATEGY 2.A ---
\multirow{4}{*}{\shortstack[l]{\textbf{Strategy 2.A} \\ \footnotesize{(Factorized} \\ \footnotesize{Prototypes)}}}
 & 0.0  & 94.93 \scriptsize{$\pm$ 0.08} & 70.94 \scriptsize{$\pm$ 0.40} & 81.44 \scriptsize{$\pm$ 0.38} & 83.97 \scriptsize{$\pm$ 0.38} & 90.05 \scriptsize{$\pm$ 0.20} & 80.86 \scriptsize{$\pm$ 0.57} & 9.19 \scriptsize{$\pm$ 0.60} \\
 & 0.05 & 95.17 \scriptsize{$\pm$ 0.13} & 71.20 \scriptsize{$\pm$ 0.29} & 81.10 \scriptsize{$\pm$ 0.11} & 83.67 \scriptsize{$\pm$ 0.15} & 91.34 \scriptsize{$\pm$ 0.14} & 79.75 \scriptsize{$\pm$ 0.22} & 11.59 \scriptsize{$\pm$ 0.26} \\
 & 0.1  & 95.21 \scriptsize{$\pm$ 0.09} & 71.03 \scriptsize{$\pm$ 0.04} & 80.81 \scriptsize{$\pm$ 0.11} & 83.33 \scriptsize{$\pm$ 0.23} & 91.51 \scriptsize{$\pm$ 0.04} & 79.14 \scriptsize{$\pm$ 0.34} & 12.37 \scriptsize{$\pm$ 0.34} \\
 & 1.0  & 95.36 \scriptsize{$\pm$ 0.18} & 68.01 \scriptsize{$\pm$ 0.83} & 78.06 \scriptsize{$\pm$ 0.66} & 80.81 \scriptsize{$\pm$ 0.56} & \textbf{91.76 \scriptsize{$\pm$ 0.08}} & 75.21 \scriptsize{$\pm$ 0.85} & 16.55 \scriptsize{$\pm$ 0.85} \\

\midrule

% --- STRATEGY 2.B ---
\multirow{4}{*}{\shortstack[l]{\textbf{Strategy 2.B}\\\footnotesize{(Residual}\\ \footnotesize{Modeling)}}}
 & 0.0  & 94.51 \scriptsize{$\pm$ 0.09} & 69.75 \scriptsize{$\pm$ 0.49} & 80.90 \scriptsize{$\pm$ 0.20} & 83.64 \scriptsize{$\pm$ 0.25} & 89.43 \scriptsize{$\pm$ 0.12} & 80.68 \scriptsize{$\pm$ 0.38} & \textbf{8.75 \scriptsize{$\pm$ 0.40}} \\
 & 0.05 & 94.62 \scriptsize{$\pm$ 0.34} & \textbf{72.11 \scriptsize{$\pm$ 0.43}} & \textbf{82.12 \scriptsize{$\pm$ 0.14}} & \textbf{84.52 \scriptsize{$\pm$ 0.22}} & 90.60 \scriptsize{$\pm$ 0.28} & \textbf{81.41 \scriptsize{$\pm$ 0.30}} &9.19 \scriptsize{$\pm$ 0.41} \\
 & 0.1  & 94.99 \scriptsize{$\pm$ 0.13} & 71.92 \scriptsize{$\pm$ 0.08} & 81.75 \scriptsize{$\pm$ 0.11} & 84.21 \scriptsize{$\pm$ 0.17} & 91.26 \scriptsize{$\pm$ 0.06} & 80.61 \scriptsize{$\pm$ 0.25} & 10.65 \scriptsize{$\pm$ 0.26} \\
 & 1.0  & 95.09 \scriptsize{$\pm$ 0.29} & 69.79 \scriptsize{$\pm$ 0.87} & 79.79 \scriptsize{$\pm$ 0.61} & 82.45 \scriptsize{$\pm$ 0.52} & 91.74 \scriptsize{$\pm$ 0.26} & 77.69 \scriptsize{$\pm$ 0.78} & 14.05 \scriptsize{$\pm$ 0.82} \\

\bottomrule
\end{tabular}
\end{table*}

\subsection{Evaluation Protocols and Scope}

We address source attribution, defining a \textbf{target class} as a specific architecture-dataset pairing (or bonafide as another source). Due to class imbalance, we report \textbf{source-level F1-Macro} as our primary metric to treat all sources equally, evaluating under two settings:

\textbf{Closed-Set Attribution (IID):} The model classifies queries ($x_q$) via cosine similarity against fixed training prototypes.

\textbf{Few-Shot Attribution Protocol:} For all sources, we compute dynamic prototypes by averaging K support embeddings per class ($K \in \{1, 3, 5\}$), then assign queries to the nearest prototype via cosine similarity.  Each configuration is evaluated over 100 few-shot trials per $K$ value and seed.

We evaluate all proposed strategies across four levels of MSE regularization strength $\lambda \in \{0.0, 0.05, 0.1, 1.0\}$. Each configuration is trained across 4 seeds.% 

\section{Results and Analysis}
\label{sec:experiments}
\label{sec:results}

We evaluate our framework under closed-set (IID) and few-shot open-set protocols (Table~\ref{tab:allkshot_f1_macro_cosine}, reports F1-Macro (\%)). To further analyze improvements at $K=5$, we decompose results into Seen vs. All and Unseen vs. All, reporting the \textbf{Generalization Gap}, the difference in performance between seen and unseen sources, to quantify degradation on novel sources.

\vspace{0.08cm}
\noindent\textbf{Baseline (ArcFace).} ArcFace achieves 95.11\% F1-Macro in closed-set, confirming its effectiveness for seen sources. However, open-set performance drops significantly (80.11\% All vs All at K=5). Critically, $K=5$ decomposition reveals a 13.70 pp generalization gap. This suggests that embeddings learned under angular margin supervision, which treats sources as monolithic class indices, do not generalize adequately to novel sources without additional geometric constraints.

\vspace{0.07cm}
\noindent\textbf{Strategy 1: Orthonormal Prototypes.} Replacing the learnable ArcFace centers with orthogonally initialized prototypes targets this open-set limitation. Strategy~1 maintains competitive Closed-Set performance (95.20--95.44\%) while consistently improving Few-Shot Open-Set attribution across all support sizes over the baseline (+2.6~pp at K=5, $\lambda=0.0$). In our framework, prototypes remain frozen after initialization; experiments with trainable prototypes yielded similar performance, supporting the fixed configuration as the more parsimonious choice. Regarding MSE regularization, a trade-off emerges: higher $\lambda$  tightly anchors embeddings, improving Seen performance (up to 91.39\%), but degrading Unseen accuracy (down to 76.35\%). This indicates that excessive geometric rigidity over-constrains the latent space, limiting adaptation to novel sources.

\vspace{0.07cm}
\noindent\textbf{Strategy 2.A: Factorized Prototypes.} Table~\ref{tab:allkshot_f1_macro_cosine} shows that Strategy~2.A ($\lambda=0.0$) improves All vs All performance over Strategy~1 (+1.3~pp at $K=5$) and achieves a substantially reduced generalization gap (9.19~pp vs. 12.07~pp for Strategy 1) with consistent Few-Shot Open-Set improvements across all support sizes. As with Strategy~1, higher $\lambda$ values progressively increase the generalization gap (up to 16.55~pp at $\lambda=1.0$), confirming that excessive geometric rigidity penalizes unseen sources regardless of the partitioning strategy.

\vspace{0.07cm}
\noindent\textbf{Strategy 2.B: Residual Modeling.} Incorporating the residual subspace $\mathcal{Z}_{\mathcal{R}}$ yields the strongest overall results: Strategy~2.B ($\lambda=0.05$) achieves the best All vs All F1-Macro at every support size (+0.55~pp over Strategy 2.A at $K=5$, reaching 84.52\%) and the highest Unseen vs All score.

\begin{figure}[!h]
    \centering
    \includegraphics[width=\columnwidth]{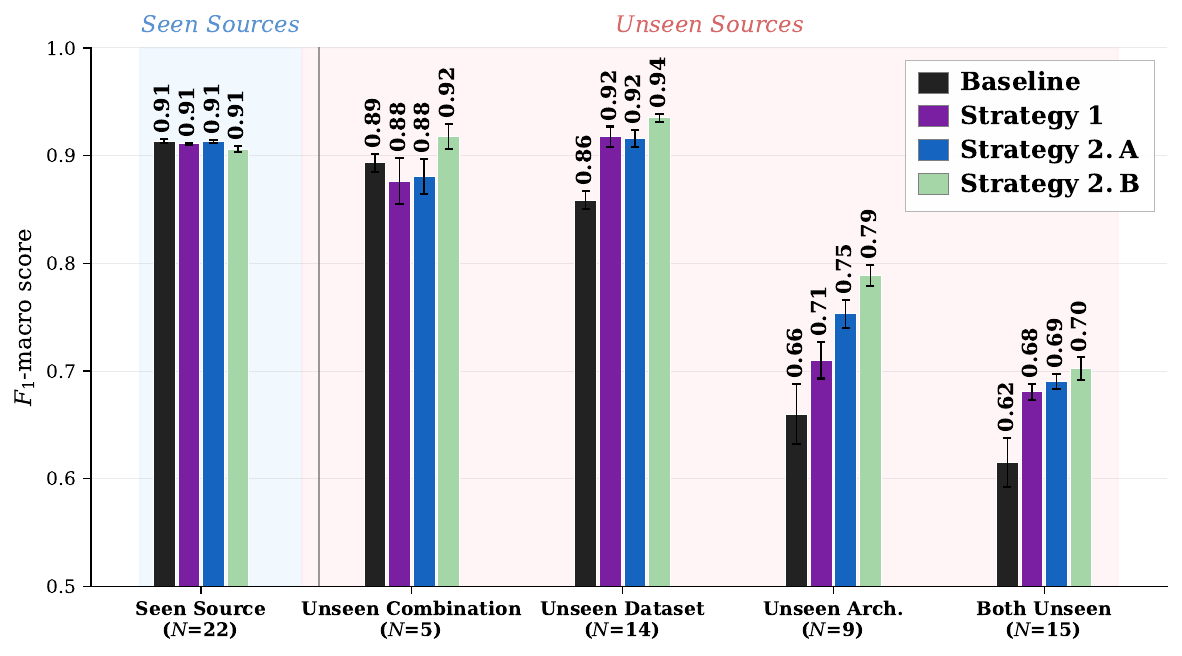}
    %\vspace{-0.6cm}
    \caption{F1-macro scores for few-shot source attribution with K= 5 support examples across five generalization settings. Best-performing configuration selected for each strategy based on validation performance. N indicates the number of test sources.}
   \label{fig:f1_macro_allk}
\end{figure}

\paragraph*{On the origin of the differences.} To better understand the gains on unseen sources, Figure~\ref{fig:f1_macro_allk} provides a per-category breakdown across the five evaluation scenarios. Strategy~1 yields significant improvements over the baseline when at least one generative factor is unseen, with gains across most OOD settings but a degradation in \textit{Unseen Combination}; the monolithic embedding cannot separate factor contributions despite the more stable geometric prior. Moreover, by forcing all sources to be mutually orthogonal, Strategy~1 erases the shared structure between pairings that reuse an architecture or a dataset, the very similarity a learnable ArcFace center can still retain, so a novel $(a,d)$ pairing has no neighboring prototype to exploit.

Introducing subspace partitioning in Strategy~2.A leaves \textit{Unseen Combination} and \textit{Unseen Dataset} largely unchanged with respect to Strategy~1. In contrast, the contribution is evident in \textit{Unseen Architecture}, suggesting that isolating $\mathcal{Z}_{\mathcal{A}}$ yields a more transferable architectural representation. Finally, gains in \textit{Both Unseen} remain modest. 

The residual subspace $\mathcal{Z}_{\mathcal{R}}$ in Strategy~2.B produces the clearest improvements in challenging settings: in \textit{Unseen Combination}, performance recovers and surpasses the baseline, suggesting that $\mathcal{Z}_{\mathcal{R}}$ captures inter-factor interactions.
The largest absolute result is achieved in \textit{Unseen Dataset}, while the most significant relative gain appears in \textit{Unseen Architecture}, indicating that known data-driven artifacts partially compensate for novel architectures. Finally, \textit{Both Unseen} gains remain limited.

Notably, the asymmetric generalization across all strategies between \textit{Unseen Dataset} and \textit{Unseen Architectures} settings suggests that architectural fingerprints act as more robust anchors within the latent space than data-driven artifacts.

\section{Conclusions}

In this work, we have redefined synthetic speech attribution as a \textbf{compositional task}, moving beyond the constraints of monolithic classification. Decomposing the generative identity into architecture ($\mathcal{A}$), training data ($\mathcal{D}$), and residual configuration ($\mathcal{H}$) factors enables generalization to novel sources through the alignment of factorized marginal representations.

Our empirical results demonstrate that while traditional angular margin losses like ArcFace excel at closed-set discrimination, they lack the \textbf{representative stability} required for robust open-set generalization. Structured orthonormal prototypes create a more consistent latent space, significantly reducing the generalization gap in few-shot scenarios. Introducing subspace partitioning further improves generalization by isolating architectural and data-driven representations, enabling better transfer when at least one generative factor has been observed during training. Finally, the addition of a dedicated residual subspace $\mathcal{Z}_{\mathcal{R}}$ acts as a critical extra manifold, absorbing non-linear interactions without compromising the primary factor representations in $\mathcal{Z}_{\mathcal{A}}$ and $\mathcal{Z}_{\mathcal{D}}$, yielding the strongest overall generalization across all OOD scenarios.

Beyond improved accuracy, the factorial structure suggests a potential path toward more \textbf{interpretable forensic analysis}: because architecture and data are aligned to separate subspaces, comparing similarity within $\mathcal{Z}_{\mathcal{A}}$ and $\mathcal{Z}_{\mathcal{D}}$ could in principle indicate whether an attribution is driven more by architectural or by data-driven cues, a direction we leave to future quantitative study. Furthermore, the compositional nature of the latent space allows the system to scale to the rapidly evolving landscape of generative AI, facilitating the incorporation of novel models and datasets by combining existing factor prototypes without the need for exhaustive retraining. This composition also scales favorably: whereas Strategy~1 requires $C\le M$ mutually orthogonal prototypes, the factorized prototypes of Strategy~2 represent up to $C_{\mathcal{A}}\times C_{\mathcal{D}}$ sources from only $C_{\mathcal{A}}\le M_{\mathcal{A}}$ and $C_{\mathcal{D}}\le M_{\mathcal{D}}$ basis vectors, a growing advantage as the number of generative models increases.
A natural next step concerns the inference variables $\mathcal{I}$ that render each utterance, such as the reference speaker. The present work does not take this variability into account and currently leaves it, together with the training residual, to be \emph{absorbed} into $\mathcal{Z}_{\mathcal{R}}$. In future work we intend to study it explicitly and \emph{compensate} for it rather than absorb it.

\section{Acknowledgments}

The authors used generative AI tools (Claude, Anthropic; Gemini, Google)
solely for language editing and polishing of the manuscript. No AI-generated material was used in the experimental design, analysis, or reporting of results, for which the authors take full responsibility.

This work has received funding from MCIN/AEI/10.13039/501100011033 under Grant PID2024-155948OB-C53.

\bibliographystyle{IEEEtran}
\bibliography{mybib}

@inproceedings{falez25_interspeech,
  title     = {{Audio Deepfake Source Tracing using Multi-Attribute Open-Set Identification and Verification}},
  author    = {Pierre Falez and Tony Marteau and Damien Lolive and Arnaud Delhay},
  year      = {2025},
  booktitle = {{Interspeech 2025}},
  pages     = {1528--1532},
  doi       = {10.21437/Interspeech.2025-2001},
  issn      = {2958-1796},
}

@inproceedings{kulkarni25_interspeech,
  title     = {{Unveiling Audio Deepfake Origins: A Deep Metric learning And Conformer Network Approach With Ensemble Fusion}},
  author    = {Ajinkya Kulkarni and Sandipana Dowerah and Tanel Alumäe and Mathew Magimai Doss},
  year      = {2025},
  booktitle = {{Interspeech 2025}},
  pages     = {1533--1537},
  doi       = {10.21437/Interspeech.2025-2079},
  issn      = {2958-1796},
}

@inproceedings{chen25j_interspeech,
  title     = {{Codec-Based Deepfake Source Tracing via Neural Audio Codec Taxonomy}},
  author    = {Xuanjun Chen and I-Ming Lin and Lin Zhang and Jiawei Du and Haibin Wu and Hung-yi Lee and Jyh-Shing Roger Jang},
  year      = {2025},
  booktitle = {{Interspeech 2025}},
  pages     = {1538--1542},
  doi       = {10.21437/Interspeech.2025-1297},
  issn      = {2958-1796},
}

@inproceedings{negroni25_interspeech,
  title     = {{ Source Verification for Speech Deepfakes }},
  author    = {Viola Negroni and Davide Salvi and Paolo Bestagini and Stefano Tubaro},
  year      = {2025},
  booktitle = {{Interspeech 2025}},
  pages     = {1548--1552},
  doi       = {10.21437/Interspeech.2025-1490},
  issn      = {2958-1796},
}

@inproceedings{dao2026speaker,
  title     = {Assessing the Impact of Speaker Identity in Speech Spoofing Detection},
  author    = {Dao, Anh-Tuan and others},
  year      = {2026},
  booktitle = {{ICASSP 2026 - IEEE International Conference on Acoustics, Speech and Signal Processing (ICASSP)}},
  organization = {IEEE},
}

@inproceedings{koutsianos25_interspeech,
  title     = {{Synthetic Speech Source Tracing using Metric Learning}},
  author    = {Dimitrios Koutsianos and Stavros Zacharopoulos and Yannis Panagakis and Themos Stafylakis},
  year      = {2025},
  booktitle = {{Interspeech 2025}},
  pages     = {1558--1562},
  doi       = {10.21437/Interspeech.2025-1757},
  issn      = {2958-1796},
}

@inproceedings{klein25_interspeech,
  title     = {{Open-Set Source Tracing of Audio Deepfake Systems}},
  author    = {Nicholas Klein and Hemlata Tak and Elie Khoury},
  year      = {2025},
  booktitle = {{Interspeech 2025}},
  pages     = {1578--1582},
  doi       = {10.21437/Interspeech.2025-1269},
  issn      = {2958-1796},
}

@inproceedings{klein24_interspeech,
  title     = {{Source Tracing of Audio Deepfake Systems}},
  author    = {Nicholas Klein and Tianxiang Chen and Hemlata Tak and Ricardo Casal and Elie Khoury},
  year      = {2024},
  booktitle = {{Interspeech 2024}},
  pages     = {1100--1104},
  doi       = {10.21437/Interspeech.2024-1283},
  issn      = {2958-1796},
}

@article{chen2025codecfake+,
  title={CodecFake+: A Large-Scale Neural Audio Codec-Based Deepfake Speech Dataset},
  author={Chen, Xuanjun and Du, Jiawei and Wu, Haibin and Zhang, Lin and Lin, I and Chiu, I and Ren, Wenze and Tseng, Yuan and Tsao, Yu and Jang, Jyh-Shing Roger and others},
  journal={arXiv preprint arXiv:2501.08238},
  year={2025}
}

@INPROCEEDINGS{10650962,
  author={Müller, Nicolas M. and Kawa, Piotr and Choong, Wei Herng and Casanova, Edresson and Gölge, Eren and Müller, Thorsten and Syga, Piotr and Sperl, Philip and Böttinger, Konstantin},
  booktitle={2024 International Joint Conference on Neural Networks (IJCNN)}, 
  title={MLAAD: The Multi-Language Audio Anti-Spoofing Dataset}, 
  year={2024},
  volume={},
  number={},
  pages={1-7},
  doi={10.1109/IJCNN60899.2024.10650962}}

@inproceedings{firc25_interspeech,
  title     = {{STOPA: A Dataset of Systematic VariaTion Of DeePfake Audio for Open-Set Source Tracing and Attribution}},
  author    = {Anton Firc and Manasi Chhibber and Jagabandhu Mishra and Vishwanath {Pratap Singh} and Tomi Kinnunen and Kamil Malinka},
  year      = {2025},
  booktitle = {{Interspeech 2025}},
  pages     = {1553--1557},
  doi       = {10.21437/Interspeech.2025-2065},
  issn      = {2958-1796},
}

@article{Deng_2022,
   title={ArcFace: Additive Angular Margin Loss for Deep Face Recognition},
   volume={44},
   ISSN={1939-3539},
   url={http://dx.doi.org/10.1109/TPAMI.2021.3087709},
   DOI={10.1109/tpami.2021.3087709},
   number={10},
   journal={IEEE Transactions on Pattern Analysis and Machine Intelligence},
   publisher={Institute of Electrical and Electronics Engineers (IEEE)},
   author={Deng, Jiankang and Guo, Jia and Yang, Jing and Xue, Niannan and Kotsia, Irene and Zafeiriou, Stefanos},
   year={2022},
   month=oct, pages={5962–5979} }

@inproceedings{almudevar24_interspeech,
  title     = {{Predefined Prototypes for Intra-Class Separation and Disentanglement}},
  author    = {Antonio Almudévar and Théo Mariotte and Alfonso Ortega and Marie Tahon and Luis Vicente and Antonio Miguel and Eduardo Lleida},
  year      = {2024},
  booktitle = {{Interspeech 2024}},
  pages     = {3809--3813},
  doi       = {10.21437/Interspeech.2024-825},
  issn      = {2958-1796},
}

@Article{electronics10070850,
AUTHOR = {Zinemanas, Pablo and Rocamora, Martín and Miron, Marius and Font, Frederic and Serra, Xavier},
TITLE = {An Interpretable Deep Learning Model for Automatic Sound Classification},
JOURNAL = {Electronics},
VOLUME = {10},
YEAR = {2021},
NUMBER = {7},
ARTICLE-NUMBER = {850},
URL = {https://www.mdpi.com/2079-9292/10/7/850},
ISSN = {2079-9292},
DOI = {10.3390/electronics10070850}
}

@inproceedings{ASVspoof15,
  title     = {ASVspoof 2015: the first automatic speaker verification spoofing and countermeasures challenge},
  author    = {Zhizheng Wu and Tomi Kinnunen and Nicholas Evans and Junichi Yamagishi and Cemal Hanilçi and Md. Sahidullah and Aleksandr Sizov},
  year      = {2015},
  booktitle = {Interspeech 2015},
  pages     = {2037--2041},
  doi       = {10.21437/Interspeech.2015-462},
  issn      = {2958-1796},
}

@inproceedings{ASVspoof17,
  title     = {The ASVspoof 2017 Challenge: Assessing the Limits of Replay Spoofing Attack Detection},
  author    = {Tomi Kinnunen and Md. Sahidullah and Héctor Delgado and Massimiliano Todisco and Nicholas Evans and Junichi Yamagishi and Kong Aik Lee},
  year      = {2017},
  booktitle = {Interspeech 2017},
  pages     = {2--6},
  doi       = {10.21437/Interspeech.2017-1111},
  issn      = {2958-1796},
}

@inproceedings{ASVspoof19,
  title     = {ASVspoof 2019: Future Horizons in Spoofed and Fake Audio Detection},
  author    = {Massimiliano Todisco and Xin Wang and Ville Vestman and Md. Sahidullah and Héctor Delgado and Andreas Nautsch and Junichi Yamagishi and Nicholas Evans and Tomi H. Kinnunen and Kong Aik Lee},
  year      = {2019},
  booktitle = {Interspeech 2019},
  pages     = {1008--1012},
  doi       = {10.21437/Interspeech.2019-2249},
  issn      = {2958-1796},
}

@ARTICLE{ASVspoof21,
  author={Liu, Xuechen and Wang, Xin and Sahidullah, Md and Patino, Jose and Delgado, Héctor and Kinnunen, Tomi and Todisco, Massimiliano and Yamagishi, Junichi and Evans, Nicholas and Nautsch, Andreas and Lee, Kong Aik},
  journal={IEEE/ACM Transactions on Audio, Speech, and Language Processing}, 
  title={ASVspoof 2021: Towards Spoofed and Deepfake Speech Detection in the Wild}, 
  year={2023},
  volume={31},
  number={},
  pages={2507-2522},
  doi={10.1109/TASLP.2023.3285283}}

@misc{ASVspoof24,
      title={ASVspoof 5: Crowdsourced Speech Data, Deepfakes, and Adversarial Attacks at Scale}, 
      author={Xin Wang and Hector Delgado and Hemlata Tak and Jee-weon Jung and Hye-jin Shim and Massimiliano Todisco and Ivan Kukanov and Xuechen Liu and Md Sahidullah and Tomi Kinnunen and Nicholas Evans and Kong Aik Lee and Junichi Yamagishi},
      year={2024},
      eprint={2408.08739},
      archivePrefix={arXiv},
      primaryClass={eess.AS},
      url={https://arxiv.org/abs/2408.08739}, 
}

@inproceedings{ADD2022,
  title        = {ADD 2022: the First Audio Deep Synthesis Detection Challenge},
  author       = {Yi, Jiangyan and Fu, Ruibo and Tao, Jianhua and Nie, Shuai and Ma, Haoxin and Wang, Chenglong and Wang, Tao and Tian, Zhengkun and Bai, Ye and Fan, Cunhang and Liang, Shan and Wang, Shiming and Zhang, Shuai and Yan, Xinrui and Xu, Le and Wen, Zhengqi and Li, Haizhou and Lian, Zheng and Liu, Bin},
  booktitle    = {ICASSP 2022 – 2022 IEEE International Conference on Acoustics, Speech and Signal Processing},
  pages        = {9216--9220},
  year         = {2022},
  doi          = {10.1109/ICASSP43922.2022.9747961},
}

@inproceedings{ADD2023,
  title        = {ADD 2023: the Second Audio Deepfake Detection Challenge},
  author       = {Yi, Jiangyan and Tao, Jianhua and Fu, Ruibo and Yan, Xinrui and Wang, Chenglong and Wang, Tao and Zhang, Chu Yuan and Zhang, Xiaohui and Zhao, Yan and Ren, Yong and Xu, Le and Zhou, Junzuo and Gu, Hao and Wen, Zhengqi and Liang, Shan and Lian, Zheng and Nie, Shuai and Li, Haizhou},
  booktitle    = {DADA@IJCAI 2023 (Second Audio Deepfake Detection Challenge Workshop)},
  pages        = {125--130},
  year         = {2023},
  arxiv        = {2408.04967},
}

@article{7b3f0a2f-4c9b-34db-b425-bd05e12a6d9b,
 ISSN = {00093920, 14678624},
 URL = {http://www.jstor.org/stable/1129703},
 author = {Sandra Scarr and Kathleen McCartney},
 journal = {Child Development},
 number = {2},
 pages = {424--435},
 publisher = {[Wiley, Society for Research in Child Development]},
 title = {How People Make Their Own Environments: A Theory of Genotype → Environment Effects},
 urldate = {2025-11-28},
 volume = {54},
 year = {1983}
}

@misc{he2015deepresiduallearningimage,
      title={Deep Residual Learning for Image Recognition}, 
      author={Kaiming He and Xiangyu Zhang and Shaoqing Ren and Jian Sun},
      year={2015},
      eprint={1512.03385},
      archivePrefix={arXiv},
      primaryClass={cs.CV},
      url={https://arxiv.org/abs/1512.03385}, 
}

@article{Pratap2020MLSAL,
  title={MLS: A Large-Scale Multilingual Dataset for Speech Research},
  author={Vineel Pratap and Qiantong Xu and Anuroop Sriram and Gabriel Synnaeve and Ronan Collobert},
  journal={ArXiv},
  year={2020},
  volume={abs/2012.03411}
}

@InProceedings{Zheng_2018_CVPR,
author = {Zheng, Yutong and Pal, Dipan K. and Savvides, Marios},
title = {Ring Loss: Convex Feature Normalization for Face Recognition},
booktitle = {Proceedings of the IEEE Conference on Computer Vision and Pattern Recognition (CVPR)},
month = {June},
year = {2018}
}

@inproceedings{bardes2022vicreg,
  title={VICReg: Variance-Invariance-Covariance Regularization for Self-Supervised Learning},
  author={Bardes, Adrien and Ponce, Jean and LeCun, Yann},
  booktitle={International Conference on Learning Representations (ICLR)},
  year={2022}
}

@article{johnson1984extensions,
  title={Extensions of Lipschitz mappings into a Hilbert space},
  author={Johnson, William B and Lindenstrauss, Joram and others},
  journal={Contemporary mathematics},
  volume={26},
  number={189-206},
  pages={1},
  year={1984}
}

@inproceedings{chhibber26_odyssey,
  title     = {{Advancing Zero-Shot Open-Set Speech Deepfake Source Tracing}},
  author    = {Manasi Chhibber and Jagabandhu Mishra and Tomi H. Kinnunen},
  year      = {2026},
  booktitle = {{Odyssey 2026}},
  pages     = {185--190},
  doi       = {10.21437/Odyssey.2026-27},
}

\end{document}